\documentclass[
prc,%
10pt,%
final,%
notitlepage,%
oneside,%
twocolumn,%
nobibnotes,%
nofootinbib,
superscriptaddress,%
floatfix,%
showkeys,%
showpacs]%
{revtex4}
\usepackage{color}
\usepackage{amsfonts}
\usepackage{amsbsy}
\usepackage{mathrsfs}
\usepackage{graphicx}
\def\lsim{\mathrel{\rlap{
\lower4pt\hbox{\hskip-3pt$\sim$}}
    \raise1pt\hbox{$<$}}}     
\def\gsim{\mathrel{\rlap{
\lower4pt\hbox{\hskip-3pt$\sim$}}
    \raise1pt\hbox{$>$}}}     

\begin{document}
\title{Robustness of the Baryon-Stopping Signal for the Onset of Deconfinement in Relativistic Heavy-Ion Collisions}%
\author{Yu. B. Ivanov}\thanks{e-mail: Y.Ivanov@gsi.de}
\affiliation{National Research Centre "Kurchatov Institute" (NRC "Kurchatov Institute"), 123182 Moscow, Russia}
\affiliation{National Research Nuclear University "MEPhI" (Moscow Engineering
Physics Institute), 115409 Moscow, Russia} 
\author{D. Blaschke}\thanks{e-mail:  blaschke@ift.uni.wroc.pl}
\affiliation{Institute of Theoretical Physics, University of Wroclaw, 50-204 Wroclaw, Poland}
\affiliation{Bogoliubov Laboratory of Theoretical Physics, JINR Dubna, 141980 Dubna, Russia}
\begin{abstract}
The impact of the experimental acceptance, i.e. transverse-momentum ($p_T$) cut-off
and limited rapidity region,
on the earlier predicted irregularity in the excitation function of the baryon stopping
is studied. 
This irregularity is a consequence of the onset of deconfinement occurring in the compression stage of a nuclear collision and manifests itself as a wiggle in the excitation function of the reduced curvature ($C_y$) of the net-proton rapidity distribution at midrapidity. 
It is demonstrated that the  wiggle is a very robust signal of a first-order phase transition that survives even under conditions of a very limited acceptance. 
At the same time the $C_y$ for pure hadronic and crossover transition scenarios become 
hardly distinguishable, 
if the acceptance cuts off  too much of the low-$p_T$ proton spectrum 
and/or puts too narrow rapidity window around midrapidity. 
It is found that the shape of the net-proton rapidity distribution near midrapidity depends on the $p_T$ 
cut-off. This implies that the measurements should be taken at the same acceptance for all collision energies in order to reliably conclude on the presence or absence of the irregularity.
\end{abstract}
\maketitle

\section{Introduction}

The onset of deconfinement in relativistic heavy-ion collisions is now in the focus of 
theoretical and experimental studies of the equation of state (EoS) and the phase diagram 
of strongly interacting matter. 
This problem is one of the main 
motivations for the currently running beam-energy scan \cite{RHIC-scan} at the 
Relativistic Heavy-Ion Collider (RHIC) at Brookhaven National Laboratory (BNL) and the
low-energy-scan program \cite{SPS-scan} at Super Proton Synchrotron (SPS)
of the European Organization for Nuclear Research (CERN)
as well as for constructing the
Facility for Antiproton and Ion Research (FAIR) in Darmstadt \cite{FAIR} and the
Nuclotron-based Ion Collider Facility (NICA) in Dubna \cite{NICA}. 

In Refs. \cite{Ivanov:2010cu,Ivanov:2011cb,Ivanov:2012bh,Ivanov:2013wha}  
it was argued that the baryon stopping 
in nuclear collision can be a sensitive probe for the onset of deconfinement. 
Rapidity distributions of net-protons were calculated \cite{Ivanov:2012bh,Ivanov:2013wha} 
within a model of the three-fluid 
dynamics (3FD) \cite{3FD} in scenarios with and without deconfinement transition. 
These calculations were performed  
employing three different types of EoS: a purely hadronic EoS   
\cite{gasEOS} (hadr. EoS) and two versions of the EoS involving    
deconfinement  \cite{Toneev06}. 
The latter two versions are an EoS with a first-order phase transition (2-phase EoS) 
and one with a smooth crossover transition (crossover EoS). 

It was found that 3FD predictions within the  first-order-transition scenario exhibit
a ``peak-dip-peak-dip'' irregularity in the 
incident energy dependence of the form of the net-proton rapidity distributions in central collisions. 
At low energies, rapidity distributions have a peak at the midrapidity. 
With the incident energy rise it transforms into a dip, then again into a peak, and with further rising
energy the midrapidity peak again changes into a dip, which already survives up to arbitrary high 
energies.
The behaviour of the type ``peak-dip-peak-dip'' in central collisions within
the 2-phase-EoS scenario is very robust with respect to variations of the model parameters in a 
wide range. 
This behaviour is in contrast with that for the hadronic-EoS scenario, 
where the form of distribution at midrapidity gradually evolves from one with a peak to one with a dip. 
The case of the crossover EoS is intermediate. Only  
a weak wiggle of the type of ``peak-dip-peak-dip'' takes place.

Experimental data also reveal a trend of the ``peak-dip-peak-dip'' irregularity 
in the energy range 8$A$ GeV $\le E_{lab}\le$ 40$A$ GeV, which is qualitatively similar 
to that in the first-order-transition scenario while quantitatively it differs. 
It is very likely that the quantitative discrepancy is due to the inadequacy of the model for
the 2-phase EoS for which the onset of the phase transition lies at rather high densities, 
above 8 times saturation density at low temperatures.
Recent models suggest an onset density of half that value  
\cite{Steinheimer:2012gc,Steinheimer:2013gla,Steinheimer:2013xxa}
which would place the onset of the wiggle structure closer to the experimentally indicated position.
However, the experimental trend is based on 
preliminary data at energies of 20$A$ GeV and  30$A$ GeV. 
Therefore, updated experimental results at energies 20$A$ and 30$A$  GeV are badly needed 
to pin down the preferable EoS and to check the hint to the wiggle behavior of the 
type ``peak-dip-peak-dip'' in the net-proton rapidity distributions. 
Moreover, 
it would be highly desirable if data in this energy range were taken 
within the same experimental setup and with the same experimental acceptance.  

The calculations of Refs.~\cite{Ivanov:2012bh,Ivanov:2013wha} were performed 
assuming the acceptance for net-protons to be wide enough to include almost all 
emitted particles. 
In practice, the extension of the acceptance beyond the range of 
$0 < p_T < 2$ GeV/c does not  practically change the net-proton rapidity distributions 
in the incident-energy range of interest. 
However, the actual experimental acceptance can be narrower. 
For the NICA MPD experiment it is restricted by the proton identification capabilities in the 
TOF detector to $0.4~{\rm GeV/c} < p_T  < 1.0~{\rm GeV/c}$ in the central rapidity range $|y|<0.5$ 
\cite{Merz:2012}.
For the case of the STAR beam energy scan the analysis of the excitation function of the
net proton rapidity distribution is still under way. In this case the acceptance will be restricted to
the range  $0.4~{\rm GeV/c} < p_T  < 3.0~{\rm GeV/c}$ and $|y|<0.5$ \cite{NuXu:2015}.

The purpose of the present paper is to investigate the question whether the ``peak-dip-peak-dip'' irregularity survives when experimental circumstances force a narrowing of the accessible acceptance
region.

\section{Equations of State} 
\label{EOS}

\begin{figure}[!hbt]
\includegraphics[width=7.0cm]{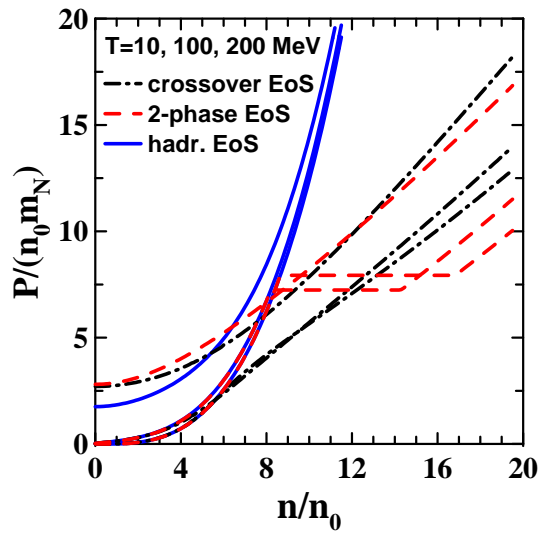}
 \caption{{
Pressure scaled by the product of normal nuclear density ($n_0=$ 0.15 fm$^{-3}$) and 
nucleon mass ($m_N$) versus baryon density scaled by the normal nuclear density
for three considered equations of state. Results are presented for three different
temperatures $T=$ 10, 100 and 200 MeV (from bottom upwards for corresponding curves).  
}} 
\label{fig1}
\end{figure}

Figure \ref{fig1} illustrates the differences between the three considered EoS. 
The deconfinement transition makes an EoS softer at high temperatures and/or densities.
The 2-phase EoS is based on the Gibbs construction, taking into account simultaneous conservation 
of baryon and strange charges. 
However, the displayed result looks very similar to the Maxwell construction, 
corresponding to the conservation of just the baryon charge, 
with the only difference that the plateau is slightly tilted, which is practically invisible. 

As demonstrated in Refs.~\cite{Ivanov:2012bh,Ivanov:2013wha}, the  
deconfinement transition in central Au+Au collisions
starts at the top AGS energies in both cases. 
It gets practically completed at low SPS energies in the case of the 2-phase EoS. 
In the crossover scenario it lasts till very high incident energies. 
We would like to note that the density range for the 2-phase EoS in Fig.~\ref{fig1} 
with an onset of deconfinement above $n\sim 8~n_0$ is rather high. 
In recent models for the spinodal decomposition accompanying the phase transition
\cite{Steinheimer:2012gc,Steinheimer:2013gla,Steinheimer:2013xxa} the onset of 
the phase coexistence is at about $4~n_0$.
This corresponds to a lower limit for the onset of the deconfinement phase transition 
in cold, symmetric nuclear matter which was obtained by a hybrid EoS model from 
constraints on the occurrence of quark matter phases in massive neutron stars
\cite{Klahn:2013kga}, see also \cite{Klahn:2011au}.

\section{
Irregularity at constrained acceptance}

The calculations at all collision energies were performed 
for Au+Au ($b=$ 2 fm) central collisions despite the fact that some  
experimental data were taken for central Pb+Pb collisions. 
This was done in order to avoid uncertainties associated with different 
colliding nuclei. 
However, in fact at the same incident energy the computed results for 
Pb+Pb collisions at $b=2.4$ fm are very close to those for Au+Au at $b=2$ fm.
The calculations were performed for four different acceptance ranges for
the transverse momentum ($p_T$) and rapidity ($y$) of the measured proton: 
\begin{enumerate}
	\item 
	$0<p_T<2$ GeV/c and a very unrestrictive constraint to the rapidity range 
	$|y|<0.7 \; y_{\rm beam}$, 
where $y_{\rm beam}$ is the beam rapidity in the collider mode, 
	 which is practically equivalent to the full acceptance; 
	\item 
	$0.4<p_T<1$ GeV/c and $|y|<0.5$, the expected MPD acceptance \cite{Merz:2012}; 
	\item 
	$1<p_T<2$ GeV/c and $|y|<0.5$, an acceptance range where low-momentum particles witnessing collective  behaviour are largely eliminated; 
	\item 
	$0.4<p_T<3$ GeV/c and $|y|<0.5$, the range of the STAR  acceptance \cite{NuXu:2015}. 
\end{enumerate}
We separately study effects of the $p_T$ and $y$ constraints in order to reveal their 
relative importance.

%
\begin{figure*}[htb]
\includegraphics[width=16.35cm]{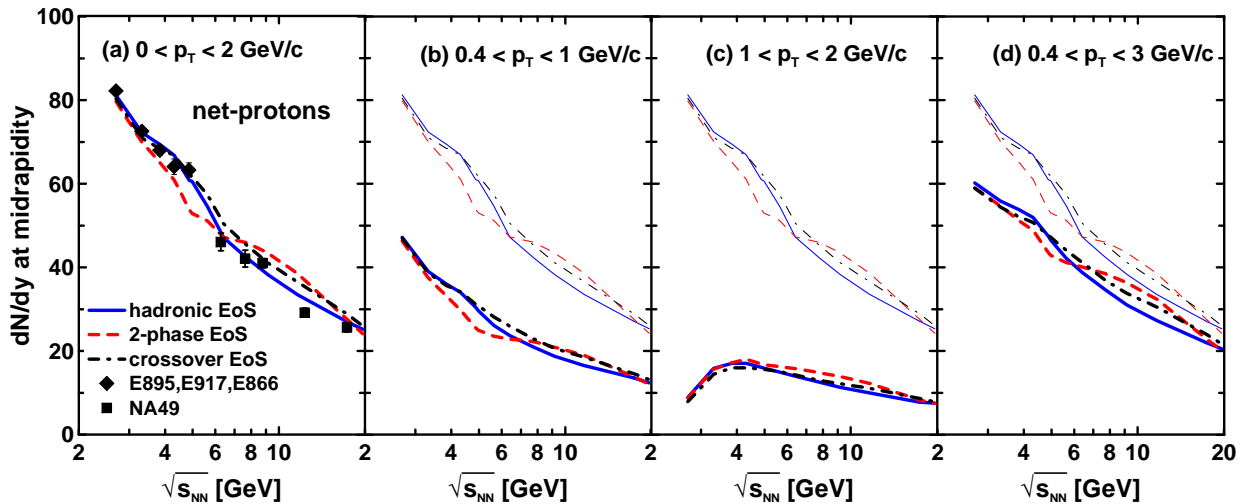}
 \caption{
Midrapidity  value 
of the net-proton rapidity distribution as a function of the collision energy 
in central ($b=$ 2 fm) Au+Au collisions in different windows of the transverse momentum $p_T$: 
 (a) $0<p_T<2$~GeV/c, (b) $0.4<p_T<1$~GeV/c, (c) $1<p_T<2$~GeV/c and d) $0.4<p_T<3$~GeV/c. 
Results for the wide $p_T$-window ($0<p_T<2$~GeV/c) are also presented in panels (b) - (d) 
by the corresponding thin lines for the sake of comparison.  
Results of simulations with different EoS's are presented.  
Experimental data are from the collaborations E895 \cite{E895}, E877 \cite{E877},
E917 \cite{E917}, E866 \cite{E866},  NA49 \cite{NA49-1,NA49-04,NA49-06,NA49-07,NA49-09} and 
STAR  \cite{STAR09}. 
}  
\label{fig2a}
\end{figure*}
%

%
\begin{figure*}[!htb]
\includegraphics[width=12.1cm]{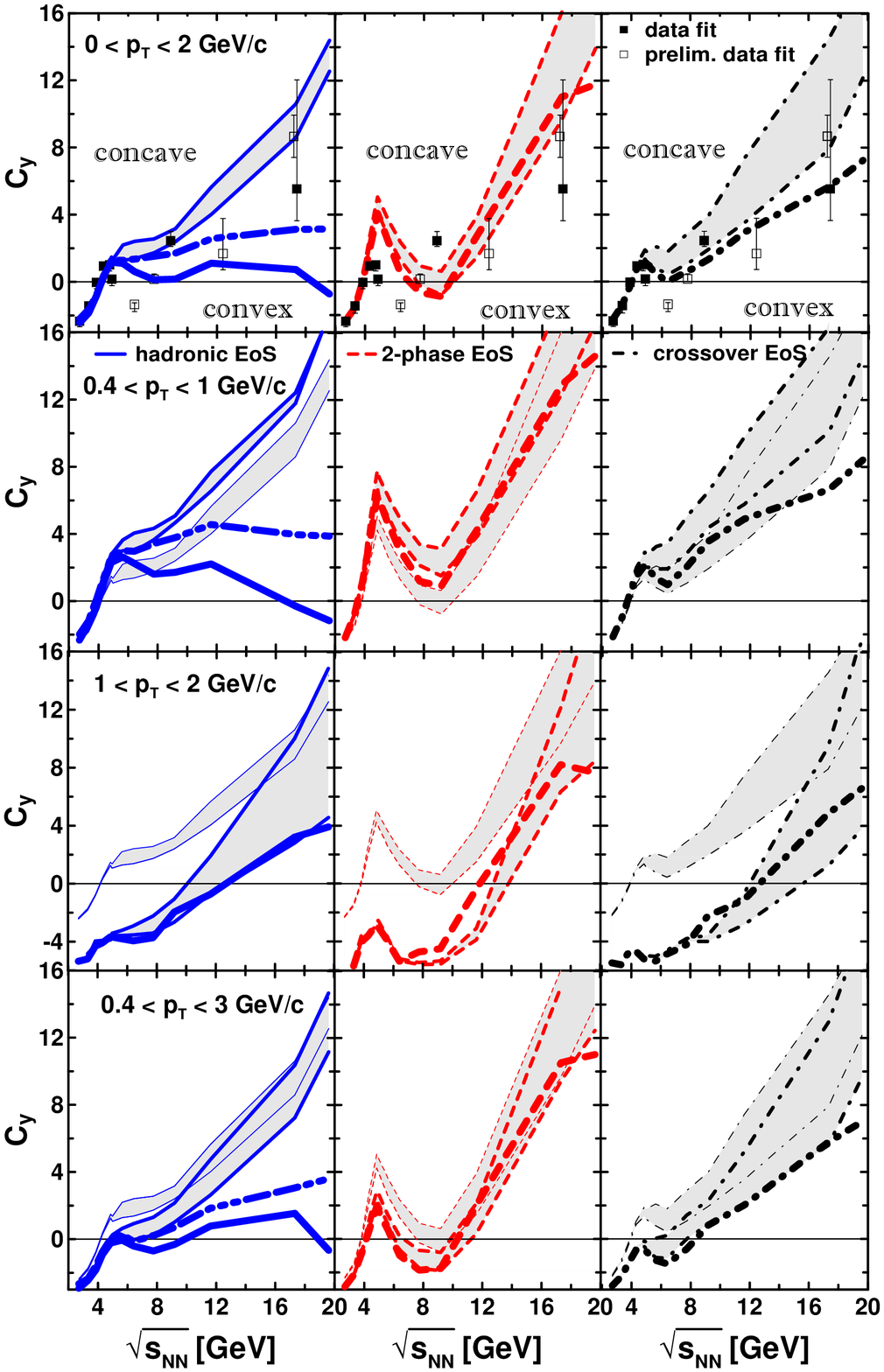}
 \caption{
Midrapidity  reduced curvature  
   [see. Eq. (\ref{Cy})] 
of the (net)proton rapidity
   spectrum as a function of the center-of-mass energy
 of colliding nuclei as deduced from experimental data and predicted
 by 3FD calculations with  different EoS's: the hadronic EoS
   (hadr. EoS) \cite{gasEOS} (left column of panels),  the EoS involving a first-order phase
 transition (2-ph. EoS, middle column of panels) and the EoS with a crossover transition 
 (crossover EoS, right column of panels) 
 into the quark-gluon phase \cite{Toneev06}. 
Upper bounds of the shaded areas correspond to fits confined in the region of 
$|y|<0.7 \; y_{\rm beam}$,  lower bounds, $|y|<0.5 \; y_{\rm beam}$.
Results are presented for 
 four different windows of the transverse momentum $p_T$: 
$0<p_T<2$~GeV/c (top row of panels), $0.4<p_T<1$~GeV/c (second row of panels), $1<p_T<2$~GeV/c (third row of panels)  and $0.4<p_T<3$~GeV/c (bottom row of panels). 
Results for the $p_T$-window of the top panel with experimental data ($0<p_T<2$~GeV) are also presented in the lower panels with different restricted $p_T$-windows 
by shaded areas bounded by the corresponding thin lines for the sake of comparison.    
Results of fits within the range of $|y|<0.5$ are displayed by corresponding bold lines. 
In those cases, when ($|y|<0.5$) fit 
substantially differs from that in the $|y|<0.9$ range, the ($|y|<0.9$)
results are also displayed by bold double-dash-triple-dotted lines. 
}
\label{fig2b}
\end{figure*}

A direct measure of the baryon stopping is the net-baryon (i.e. baryons-minus-antibaryons) rapidity distribution. 
However, since experimental information on neutrons is unavailable, we have to rely on net-proton 
(i.e. proton-minus-antiproton) data. 
Presently there exist experimental data on proton (or net-proton) rapidity spectra at AGS \cite{E895,E877,E917,E866} and SPS \cite{NA49-1,NA49-04,NA49-06,NA49-07,NA49-09} energies. 
At AGS energies, the yield of antiprotons is negligible, therefore 
the proton rapidity spectra serve as a good probe of the baryon stopping.

In order to quantify the above-discussed ``peak-dip-peak-dip'' irregularity, it is useful to make use of the method proposed in Ref.~\cite{Ivanov:2010cu}. 
For this purpose the data on the net-proton rapidity distributions are fitted  by a simple formula 
\begin{eqnarray}
\label{2-sources-fit} 
\frac{dN}{dy}&=&  
a \left(\exp\left\{ -(1/w_s)  \cosh(y-y_s) \right\}
\right.
\cr
&&+
\left.
\exp\left\{-(1/w_s)  \cosh(y+y_s)\right\} \right)~,
\end{eqnarray}
where $a$, $y_s$ and $w_s$ are parameters of the fit. 
The form (\ref{2-sources-fit}) is a sum of two thermal sources shifted by $\pm
y_s$ from the midrapidity
which is put to be $y_{\rm mid}=0$ as it is in the collider mode. 
The width $w_s$ of the sources can be interpreted as $w_s=$ (temperature)/(transverse mass), 
if we assume that collective velocities in the sources have no spread with respect
to the  source rapidities $\pm y_s$. 
The parameters of the two sources are identical (up to the sign of  $y_s$) because  only 
collisions of identical nuclei are considered. 

The above fit has been prepared using the least-squares method and applied to both, 
available data and results of calculations.  
The fit was performed in the rapidity range $|y|<0.7 \; y_{\rm beam}$, 
where $y_{\rm beam}$ is the beam rapidity in the collider mode. 
The choice of this range is dictated by the data. 
As a rule, the data are available in this rapidity range, sometimes the data range is even
more narrow (80$A$ GeV and new data at 158$A$ GeV \cite{NA49-09}). 
We apply the above restriction in order to deal with different data in
approximately the same rapidity range. 
Another reason for this cut is that the rapidity range should not be too wide in order to
exclude contributions of cold spectators.
The fit in the rapidity range 
$|y|<0.5 \; y_{\rm beam}$ 
has been also done 
in order to estimate uncertainty of the fit parameters associated with the choice of fit range.
An additional set of fits has been done under the constraint $|y|<0.5$. 
This fit was applied only to the 3FD simulation results.  
Due to low number of experimental points within the $|y|<0.5$ range and their
insufficient accuracy such a fit gives too large error bars for the deduced parameters of the fit 
(\ref{2-sources-fit}) at $\sqrt{s_{NN}}>$ 5 GeV, whereas it is practically identical to the 
$|y|<0.7 \; y_{\rm beam}$ fit at $\sqrt{s_{NN}}<$ 5 GeV. 

A useful quantity, which characterizes the shape of the rapidity distribution, is
a reduced curvature of the spectrum at midrapidity defined as follows 
\begin{eqnarray}
\label{Cy} 
C_y &=& 
\left(y_{\rm beam}^3\frac{d^3N}{dy^3}\right)_{y=0}
\big/ \left(y_{\rm beam}\frac{dN}{dy}\right)_{y=0}
\cr
&=&  
(y_{\rm beam}/w_s)^2 \left(
\sinh^2 y_s -w_s \cosh y_s 
\right). 
\end{eqnarray}
The factor $1/\left(y_{\rm beam}dN/dy\right)_{y=0}$
is introduced in order 
to cancel out the overall normalization of the spectrum. 
The second part of Eq. (\ref{Cy}) presents this curvature in terms of the parameters of 
the fit (\ref{2-sources-fit}).
The reduced curvature, $C_y$, and the midrapidity value, 
$\left(dN/dy\right)_{y=0}$, 
are two independent quantities quantifying the 
the spectrum in the midrapidity range. 
Excitation functions of these quantities deduced both from 
experimental data and from 
results of the 3FD calculations with different EoS's are displayed in Figs.~\ref{fig2a} 
and \ref{fig2b}. 

In Fig.~\ref{fig2a} the midrapidity  values of the rapidity spectra were taken directly from experimental 
data and calculated results. 
Therefore, only experimental error bars are displayed there. 
The uncertainty associated with the choice of the 
rapidity range turned out to be the dominant one for the $C_y$
quantities deduced  from both experimental data and simulation results. 
Therefore, in Fig.~\ref{fig2b} results for the curvature $C_y$
in the wide rapidity range
are presented by shaded areas with borders corresponding 
to the fit ranges $|y|<0.7 \; y_{\rm beam}$ and $|y|<0.5 \; y_{\rm beam}$. 
The $C_y$ results in the narrow rapidity range $|y|<0.5$, corresponding to 
the MPD and STAR acceptance, are also displayed by bold lines. 
In order to control the error induced by the narrow range $|y|<0.5$, 
the $C_y$ calculations were also performed in the range $|y|<0.9$. 
In most cases the ($|y|<0.5$) and ($|y|<0.9$) results turned out to be very close 
to each other. In those few cases, when ($|y|<0.5$) and ($|y|<0.9$) results
substantially differ, the ($|y|<0.9$) results are also displayed in Fig.~\ref{fig2b}.
All presently available 
experimental data shown in the leftmost panel of Fig.~\ref{fig2a}
and in the top panels of Fig.~\ref{fig2b} 
approximately correspond to acceptance range (i).

As seen from Fig.~\ref{fig2a}, approximately 60\% (for different collision energies and EoS's) 
of the protons produced at midrapidity will be covered by the MPD acceptance window,    
while the corresponding coverage of the STAR acceptance is $\sim$80\%.

The irregularity in the data  is distinctly seen  
as a strong wiggle in the excitation function of $C_y$, see Fig.~\ref{fig2b}. 
Various data used to deduce $C_y$ approximately correspond to the 
wide acceptance window (i). 
Of course, this is only a hint to an irregularity since this wiggle is formed only  
by preliminary data of the NA49 collaboration. 
In the wide acceptance window (i) (shaded bands in the upper raw of panels in Fig.~\ref{fig2b})  
the $C_y$ excitation function in the first-order-transition
scenario manifests qualitatively (though not quantitatively) the same wiggle irregularity 
(middle-column upper-raw panel in Fig. \ref{fig2b}) like
that in the data fit, while the hadronic  scenario produces purely monotonous 
behavior.  
The crossover EoS represents a very smooth transition. Therefore, 
it is not surprising that it produces only a weak wiggle in $C_y$. 

The application of various $p_T$ cuts without confining the rapidity range 
(shaded bands in Fig.~\ref{fig2b}) does not qualitatively change the picture. 
As seen from Fig.~\ref{fig2b}, the wiggle in the energy dependence of $C_y$ 
is a very robust signal of the first-order phase transition. 
It survives even at very limited $p_T$-acceptance. 
The important difference between different $p_T$-acceptances is that the wiggle 
is completely located in the range of positive curvatures $C_y$ 
(concave shapes of the rapidity distribution near midrapidity) at low-$p_T$ 
acceptance, like the MPD one (ii).  While for the high-$p_T$ 
acceptance (iii), the wiggle entirely lies in the range of negative $C_y$
(convex shape).  
Therefore, the name of "peak-dip-peak-dip" for this irregularity becomes 
not quite correct.
The amplitude of the wiggle becomes somewhat weaker when the  
$p_T$ cut-off from below is too strong, e.g., when $1<p_T<2$ GeV/c. It is expectable because
the wiggle is an effect of collective behaviour of the system, in which predominantly 
low-momentum particles participate. If these low-momentum particles are cut off by the acceptance, 
the collective effects, in particular, the wiggle, become less manifested. 

The $C_y$ excitation functions 
for hadronic and crossover scenarios are very similar to each other 
already in the wide acceptance window (i) (the upper raw of panels in Fig.~\ref{fig2b}). 
The crossover scenario results in a very weak wiggle in $C_y$. 
Under the MPD (ii) and STAR (iv) $p_T$ cuts the 
basic features of the $C_y$ excitation functions remain the same, 
only the weak crossover wiggle turns out to be shifted to the range 
of higher (the MPD case) or lower (the STAR case)  $C_y$. 
At the high-$p_T$ cut (iii) the qualitative difference 
between predictions of the hadronic and crossover scenarios is 
practically washed out. This a consequence of the above-mentioned lack 
of collectivity in the behavior of the high-momentum particles. 

\begin{figure}[htb]
\includegraphics[width=5.35cm]{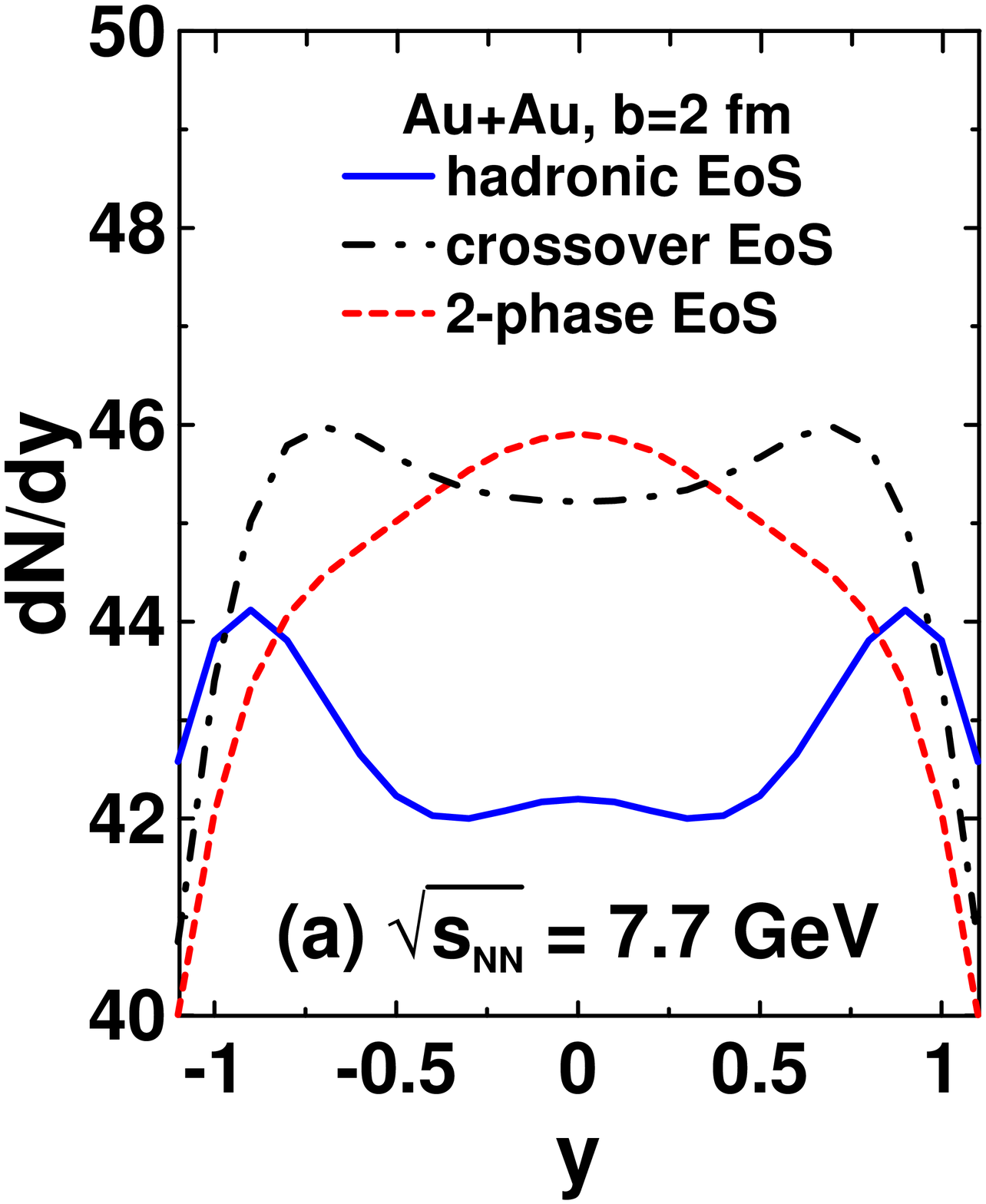}
\includegraphics[width=5.35cm]{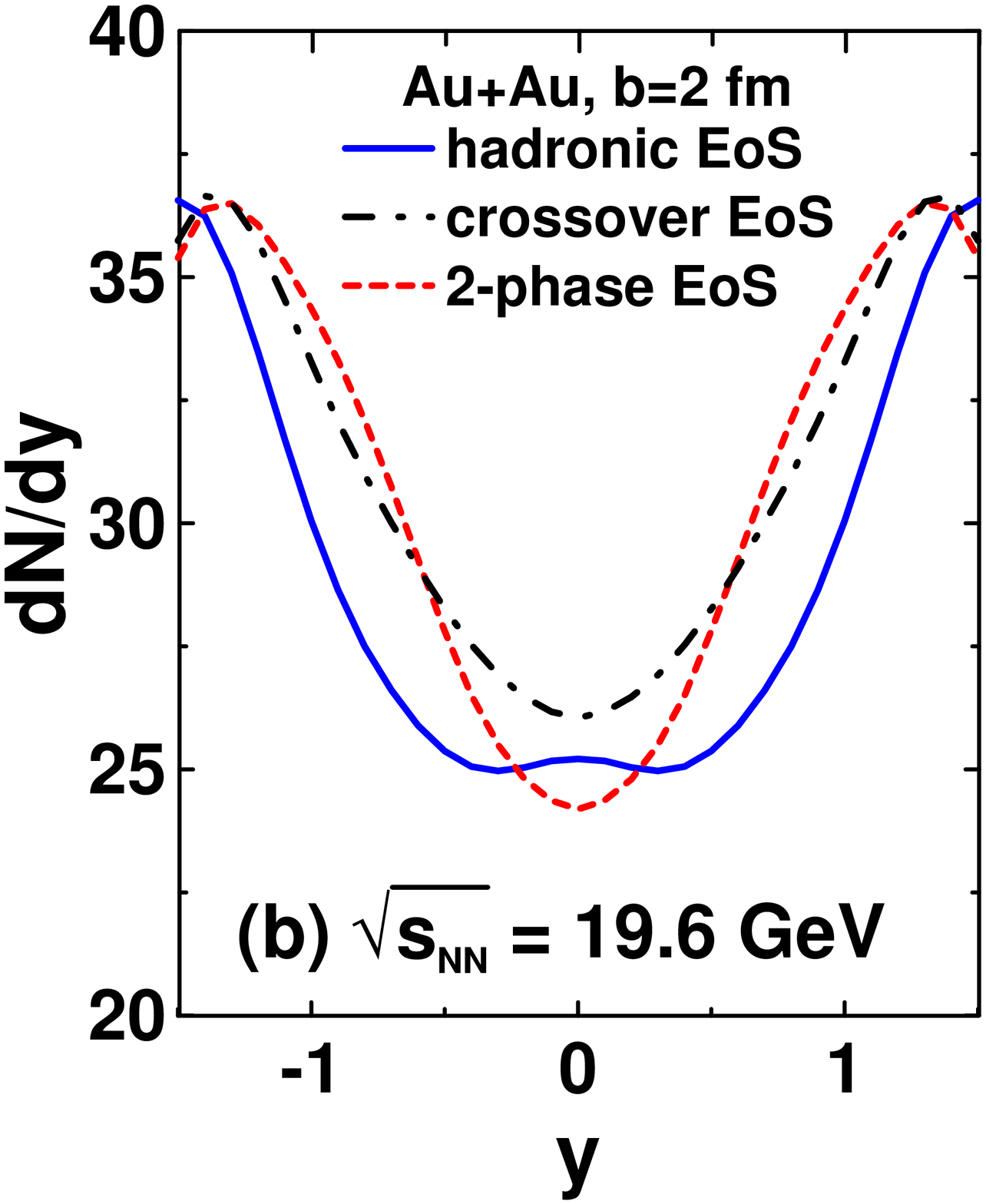}
 \caption{
Rapidity  distribution  
of net-protons in central ($b=$ 2 fm) Au+Au collisions at collision energies 
$\sqrt{s_{NN}}=$ 7.7 GeV (a) and 19.6 GeV (b) for the case of the $p_T$ range
$0<p_T<2$~GeV/c. 
}  
\label{fig3}
\end{figure}

The actual MPD (ii) and STAR (iv) acceptance conditions also include the restriction 
$|y|<0.5$. 
Applying this restriction in addition to the $p_T$ cuts
leads to the corresponding bold lines in Fig.~\ref{fig2b}. 
In order to study how restrictive the condition $|y|<0.5$ is, 
calculations were also done for the rapidity window $|y|<0.9$. 
In most cases the $C_y$ results for the range $|y|<0.5$ practically 
coincide with those in the $|y|<0.9$ window and are quite close to 
those in the wide rapidity window $|y|<0.5 \; y_{\rm beam}$. 
In these cases the $|y|<0.9$ results are not displayed in Fig.~\ref{fig2b}. 
In a few cases the $|y|<0.5$ and $|y|<0.9$ results considerably differ. 
These cases correspond to the hadronic EoS when low-$p_T$ protons are 
included in the analysis. For these cases the $|y|<0.9$ results are displayed 
by bold double-dash-dash-triple-dotted lines in Fig.~\ref{fig2b} 
(contrary to solid bold lines for the $|y|<0.5$ results).

As seen from Fig.~\ref{fig2b}, under the constraints of the MPD (ii) and STAR (iv) 
acceptance conditions the wiggle in the energy dependence of $C_y$ 
is very robust for the first-order phase transition and the crossover one. 
However, this is not the case for the hadronic EoS. 
The hadronic-EoS $C_y$ excitation functions now exhibit a local weak wiggle 
in the region 8 $<\sqrt{s_{NN}}<$ 12 GeV
(in all cases except for that with the high-$p_T$ range $1<p_T<2$ GeV/c)
qualitatively similar to that for the crossover transition. 
Moreover, the $C_y$ curvature becomes negative at $<\sqrt{s_{NN}}=$ 19.6 GeV 
(again in all cases except for that with the high-$p_T$ range). 
These peculiarities are consequences of a fine structure of the rapidity 
distribution near midrapidity that becomes dominant in the narrow rapidity 
window $|y|<0.5$. This situation is illustrated in Fig.~\ref{fig3}. 
It is seen that at $\sqrt{s_{NN}}=$ 7.7 and 19.6 GeV there are tiny maxima at the midrapidity 
in the hadronic-EoS rapidity distributions, 
which  result in the weak wiggle and negative curvature, respectively,
observed in Fig.~\ref{fig2b}.  The global shape of the hadronic-EoS rapidity distributions
does not exhibit such peculiarities, as it is clear from $C_y$ calculations in a 
wider rapidity range. 
For a comparison, see also Fig.~4 of Ref.~\cite{Mitrovski:2008hb}.

Therefore, we can conclude that under the conditions of the the MPD and STAR acceptance
it is possible to distinguish the the first-order phase transition, the onset 
of which is signalled by a strong wiggle in the excitation function of $C_y$. 
However, the difference between the purely hadronic case and that with the smooth 
crossover transition becomes ambiguous.

\section{Conclusion}

An irregularity in the baryon stopping is a natural consequence of deconfinement  
occurring in the compression stage of a nuclear collision. It is a combined effect 
of the softest point \cite{Hung:1994eq,Steinheimer:2012bp} of an EoS and a change 
in the nonequilibrium regime from hadronic to partonic one.  
As was demonstrated in Refs. \cite{Ivanov:2012bh,Ivanov:2013wha}, 
this irregularity manifests itself as a wiggle in the excitation function
of a reduced curvature ($C_y$) of the  net-proton rapidity distribution at midrapidity. 
These calculations were performed in the full acceptance range for protons. 
In the present paper we studied the effect of a restricted acceptance, in particular, 
the one expected for the NICA MPD experiment and the one of the STAR beam energy scan 
program, on the earlier predicted $C_y$ wiggle. 

It was found that the wiggle in the excitation function of $C_y$ is a quite robust signal of 
the onset of deconfinement as a first-order phase transition. 
It survives even at very limited acceptance. 
Therefore, the MPD experiment will be suitable for the experimental investigation of this effect.  
It was also found that the shape of the net-proton rapidity distribution near the
midrapidity depends on the experimental acceptance. 
Hence, only if the measurements are taken at the same acceptance for all collision energies
we can reliably conclude on the presence or absence of the wiggle. 
Notice that this is not the case for the data analysed so far. 
These data were measured in four different experiments. 
Therefore, the question if the presently available data really indicate a wiggle in the $C_y$
excitation function is still open. 
It is the more so because some of the data in the region of the expected wiggle still have a 
preliminary status. 
Data in preparation by the STAR experiment will meet the requirement that the excitation function 
of  the curvature $C_y$ shall be measured by one experiment and under same acceptance 
conditions since they stem from a collider experiment with an acceptance range widely independent 
of the beam energy.
These data, however, come from the energy range above $\sqrt{s_{NN}}=7.7$ GeV, while the wiggle is expected in the range between $4$ and $8$ GeV.
Therefore, systematic measurements with the NICA MPD collider experiment 
in the whole energy range of interest
would be highly desirable to clarify this problem. 

It is important to emphasize that the discussed irregularity 
is a signal from the hot and dense stage of the nuclear collision.  
It is formed at the (nonequilibrium) compression stage of the collision, 
as it was argued in Refs. \cite{Ivanov:2012bh,Ivanov:2013wha}. 
Therefore, a theoretical approach should be able to treat differences in 
the models for the EoS (in particular, absence or presence of a first-order 
deconfinement transition) already at this early nonequilibrium compression stage of the collision. 
The 3FD model does it by means of the three-fluid interactions. 
This, at least, makes the baryon stopping EoS dependent. 
At the same time, a hybrid fluid model does not distinguish different 
EoS models at the early collision stage because the initial state for the
hydrodynamical evolution is prepared by means of the same hadronic kinetic model for any EoS. 
Hence, the hybrid fluid model mainly reveals the baryon stopping inherent in the hadronic kinetic model
and cannot produce an irregularity of this stopping relevant to the hydrodynamical EoS. 
Here the situation is similar to the case of the directed flow that is also predominantly formed at the 
early nonequilibrium stage of the collision (see discussion in 
Ref. \cite{Ivanov:2014ioa}). 

\begin{acknowledgments} 
We are grateful to A.S. Khvorostukhin, V.V. Skokov,  and V.D. Toneev for providing 
us with the tabulated 2-phase and crossover EoS's.
Our thanks go to O. Rogachevsky and V. Voronyuk
for information about the acceptance range of the TOF detector in the NICA MPD experiment
and to Nu Xu for the case of the STAR experiment. 
D.B. acknowledges discussions with M. Bleicher, Y. Karpenko, H. Petersen and J. Steinheimer 
during his stay at FIAS Frankfurt. 
The calculations were performed at the computer cluster of GSI (Darmstadt). 
Y.B.I. received partial support from  the Russian Ministry of Science and Education 
grant NS-932.2014.2, the work of D.B.  was supported in part by the Polish NCN under grant 
UMO-2011/02/A/ST2/00306 and by the Hessian LOEWE initiative through HIC for FAIR.
\end{acknowledgments}
 


\begin{thebibliography}{999}
%
\bibitem{RHIC-scan}
G.~S.~F.~Stephans,
  J.\ Phys.\ G {\bf 32}, S447 (2006)
  [nucl-ex/0607030].
%
\bibitem{SPS-scan}
P. Seyboth [NA49 Collaboration], Addedndum-1 to the NA49 Proposal, CERNSPSC-
97-26;
M. Gazdzicki, nucl-th/9701050; 
M. Gazdzicki et al. [NA61/SHINE Collaboration], PoS C POD2006, 016 (2006).
%
\bibitem{FAIR}  
  B.~Friman, (ed.), C.~Hohne, (ed.), J.~Knoll, (ed.), S.~Leupold, (ed.), J.~Randrup, (ed.), R.~Rapp, (ed.) and P.~Senger, (ed.),
  Lect.\ Notes Phys.\  {\bf 814}, 1 (2011).
%
\bibitem{NICA} 
A.~N.~Sissakian, A.~S.~Sorin and V.~D.~Toneev,
  Conf.\ Proc.\ C {\bf 060726}, 421 (2006)
  [nucl-th/0608032].
%
%
\bibitem{Ivanov:2010cu}
  Yu.~B.~Ivanov,
  Phys.\ Lett.\  B {\bf 690}, 358 (2010)
  [arXiv:1001.0670 [nucl-th]].
%
\bibitem{Ivanov:2011cb} 
  Yu.~B.~Ivanov,
Phys. At. Nucl. {\bf 75} 621 (2012)  
  [1101.2092 [nucl-th]].
%
\bibitem{Ivanov:2012bh} 
  Yu.~B.~Ivanov,
  Phys.\ Lett.\ B {\bf 721}, 123 (2013)
  [arXiv:1211.2579 [hep-ph]].
%
\bibitem{Ivanov:2013wha} 
  Yu.~B.~Ivanov,
  Phys.\ Rev.\ C {\bf 87}, no. 6, 064904 (2013)
  [arXiv:1302.5766 [nucl-th]].
%
\bibitem{3FD}
 Yu. B. Ivanov, V. N. Russkikh, and V.D. Toneev,
 Phys. Rev. C {\bf 73}, 044904 (2006) [nucl-th/0503088].
%
\bibitem{gasEOS}
V. M. Galitsky and I. N. Mishustin, Sov. J. Nucl. Phys. {\bf 29}, 181
(1979).
%
\bibitem{Toneev06}
A. S. Khvorostukhin,  
V. V. Skokov, K. Redlich, and V. D. Toneev,
Eur. Phys. J. {\bf C48}, 531 (2006) [nucl-th/0605069].

\bibitem{Steinheimer:2012gc} 
  J.~Steinheimer and J.~Randrup,
  Phys.\ Rev.\ Lett.\  {\bf 109}, 212301 (2012)
  [arXiv:1209.2462 [nucl-th]].
\bibitem{Steinheimer:2013gla} 
  J.~Steinheimer and J.~Randrup,
  Phys.\ Rev.\ C {\bf 87}, no. 5, 054903 (2013)
  [arXiv:1302.2956 [nucl-th]].
\bibitem{Steinheimer:2013xxa} 
  J.~Steinheimer, J.~Randrup and V.~Koch,
  Phys.\ Rev.\ C {\bf 89}, no. 3, 034901 (2014)
  [arXiv:1311.0999 [nucl-th]].
\bibitem{Klahn:2013kga} 
  T.~Kl\"ahn, R.~{\L}astowiecki and D.~B.~Blaschke,
  Phys.\ Rev.\ D {\bf 88}, no. 8, 085001 (2013)
  [arXiv:1307.6996].
%
\bibitem{Klahn:2011au} 
  T.~Kl\"ahn, D.~Blaschke and F.~Weber,
  Phys.\ Part.\ Nucl.\ Lett.\  {\bf 9}, 484 (2012)
  [arXiv:1101.6061 [nucl-th]].
\bibitem{Merz:2012}
 S.~P.~Merz, S.~V.~Rasin and O.~V.~Rogachevsky,
 talk at the Russian Academy of Sciences (2012);
 {http://mpd.jinr.ru/data/presentations/ras/merts.pdf}
 \bibitem{NuXu:2015}
 N.~Xu, private communication (2015).
%
\bibitem{E895} 
  J.~L.~Klay {\it et al.}  [E-0895 Collaboration],
  Phys.\ Rev.\ C {\bf 68}, 054905 (2003)
  [nucl-ex/0306033].
%
\bibitem{E877}  J. Barrette {\em et al.} (E877 Collab.),
Phys. Rev. C {\bf 62}, 024901 (2000).
%
\bibitem{E917} B. B. Back et al., (E917 Collab.),
 Phys. Rev. Lett. {\bf 86 }, 1970 (2001).
%
\bibitem{E866} 
J. Stachel, Nucl. Phys. {\bf A654}, 119c (1999)
[nucl-ex/9903007]. 
%
\bibitem{NA49-1}  H. Appelsh\"auser {\em et al.} (NA49 Collab.),
  Phys. Rev. Lett. {\bf 82}, 2471 (1999).
%
\bibitem{NA49-04}   T. Anticic {\em et al.} (NA49 Collab.),
   Phys. Rev. C {\bf 69}, 024902 (2004).
%
\bibitem{NA49-06}  C. Alt   {\em et al.} (NA49 Collab.),
Phys. Rev. C {\bf 73}, 044910 (2006)
[nucl-ex/0512033].
%
\bibitem{NA49-07}   
C. Blume (NA49 Collab.), J. Phys. {\bf G34}, S951 (2007)
[nucl-ex/0701042].
%
\bibitem{NA49-09}   
  T.~Anticic {\it et al.}  [NA49 Collaboration],
  Phys.\ Rev.\ C {\bf 83}, 014901 (2011)
  [arXiv:1009.1747 [nucl-ex]].
%
\bibitem{STAR09}
B. I. Abelev   {\it et al.}  [STAR Collab.], 
 Phys. Rev.     C {\bf 79}, 034909  (2009) 
 [arXiv:0808.2041 [nucl-ex]]. 
%
\bibitem{Mitrovski:2008hb} 
  M.~Mitrovski, T.~Schuster, G.~Graf, H.~Petersen and M.~Bleicher,
  Phys.\ Rev.\ C {\bf 79}, 044901 (2009)
  [arXiv:0812.2041 [hep-ph]].

%
\bibitem{Hung:1994eq} 
  C.~M.~Hung and E.~V.~Shuryak,
  Phys.\ Rev.\ Lett.\  {\bf 75}, 4003 (1995)
  [hep-ph/9412360].
\bibitem{Steinheimer:2012bp} 
  J.~Steinheimer and M.~Bleicher,
  Eur.\ Phys.\ J.\ A {\bf 48}, 100 (2012)
  [arXiv:1207.2792 [nucl-th]].
%
\bibitem{Ivanov:2014ioa} 
Y.~B.~Ivanov and A.~A.~Soldatov,
  Phys.\ Rev.\ C {\bf 91}, no. 2, 024915 (2015)
  [arXiv:1412.1669 [nucl-th]].

\end{thebibliography}
\end{document}